\documentclass[desactivate]{aa}  

\usepackage{graphicx}
\usepackage[utf8]{inputenc}
\usepackage[T1]{fontenc}
\usepackage{siunitx}
\usepackage{placeins}
\usepackage{txfonts}
\usepackage{hyperref}
\hypersetup{colorlinks, linkcolor=blue, citecolor=blue}
\begin{document}
   
   \title{Kinematical signatures: Distinguishing between warps and radial flows}

   \author{A. Zuleta \inst{1},
            T. Birnstiel \inst{1, 2},
            R. Teague \inst{3}}
          
   \institute{University Observatory, Faculty of Physics, Ludwig-Maximilians-Universit\"{a}t M\"{u}nchen, Scheinerstr. 1, 81679 Munich, Germany
         \and
             Exzellenzcluster ORIGINS, Boltzmannstr. 2, D-85748 Garching, Germany
         \and   
            Department of Earth, Atmospheric, and Planetary Sciences, Massachusetts Institute of Technology, Cambridge, MA 02139, USA
             }

 
  \abstract
   {Increasing evidence shows that warped disks are common, challenging the methods used to model their velocity fields. Molecular line emission of these disks is characterized by a twisted pattern, similar to the signal from radial flows, complicating the study of warped disk kinematics. Previous attempts to model these features have encountered difficulties in distinguishing between the underlying kinematics of different disks.}
   {This study aims to advance gas kinematics modeling capabilities by extending the Extracting Disk Dynamics (\texttt{eddy}) package to include warped geometries and radial flows. We assess the performance of \texttt{eddy} in recovering input parameters for scenarios involving warps, radial flows, and combinations of the two. Additionally, we provide a basis to break the visual degeneracy between warped disks and radial flow, establishing a criterion to distinguish them.}
   {We extended the \texttt{eddy} package to handle warped geometries by including a parametric prescription of a warped disk and a ray-casting algorithm to account for the surface self-obscuration arising from the 3D to 2D projection. The effectiveness of the tool was tested using the radiative transfer code \texttt{RADMC3D}, generating synthetic models for disks with radial flows, warped disks, and warped disks with radial flows.}
   {We demonstrate the efficacy of our tool in accurately recovering the geometrical parameters of systems, particularly in data with sufficient angular resolution. Importantly, we observe minimal impact from thermal noise levels typical in Atacama Large Millimeter/submillimeter Array (ALMA) observations. Furthermore, our findings reveal that fitting an incorrect model type produces characteristic residual signatures, which serve as kinematic criteria for disk classification.}
   {Characterizing gas kinematics requires careful consideration of twisted motions. While our model provides insights into disk geometries, caution is needed when interpreting parameters in regions with complex kinematics or low-resolution data. Future ALMA baseline observations should help clarify warped disk kinematics.}


   \maketitle
%

\section{Introduction}

    Warped disks, whose orbital plane changes with radius, appear broadly in astrophysics. Theoretical predictions of this phenomenon result from a viscous disk evolving under external torques \citep{Papaloizoy1983, Pringle1992}. The ubiquity of warped disks spans diverse astronomical environments, as evidenced by observations ranging from galactic disks \citep[e.g.,][]{Sanchez2003, Chen2019}, X-ray binaries \citep[e.g.,][]{Wijers1999, Begelman2006}, active galactic nuclei \citep[e.g.,][]{Herrnstein1996, Greenhill2003}, Class 0 objects \citep[IRAS 04368+2557][]{Sakai2018}, Class 1 objects \citep[L1489 IRS][]{Sai2020}, and debris disks \citep[e.g.,][]{Mouillet1997, Currie2012, Kasper2015}. In protoplanetary disks, warps could be induced by a planet in a misaligned orbit \citep{Nealon2018}, or binary companions \citep{Facchini2018}. In the presence of magnetized stars, in which both the magnetic and rotational axes are tilted relative to the disk's rotational axis, warped disks are expected to be formed \citep{Romanova2021}. Additionally, hydrodynamical simulations suggest that misalignments between the outer and inner regions of a primordial disk may arise after late infall events \citep{Kuffmeier2021}. Warps can also be generated by stellar flybys \citep{Xiang2016, Cuello2019, Mayama2020}, which produce recognizable observable signatures for these kinds of interacting systems \citep{Cuello2020}.

    Observational signatures of warped geometries can be seen in near-infrared scattered light images as azimuthal dips in brightness \citep{Marino2015}. Likewise, nulls in scattered light observations can be produced by polarization self-cancellation \citep{Weber2023}, though this mechanism is only significant in observations with multiple illumination sources. These structures, commonly referred to as shadows, can be categorized into narrow and broad classes. Narrow shadows only extend a few degrees and result from a highly inclined inner disk. This feature was first observed, but not reported, in HD 142527 \citep{Fukagawa2006}, and later identified as intensity nulls \citep{Casassus2012}. Narrow shadows have since been reported in various systems, including HD100453 \citep{Wagner2015, Benisty2017}, HD 97048 \citep{vanderPlas2017}, HD 135344B \citep{Stolker2017}, DoAr44 \citep{Casassus2018}, GG Tau \citep{Keppler2020}, and ZZ Tau IRS \citep{Hashimoto2024}. In contrast, broad shadows obscure a significant portion of the disk and arise from a slightly inclined inner disk. These characteristics can be found in sources such as TW Hya \citep{Debes2017} and HD 1430006 \citep{Benisty2018}. A remarkable case is HD 139614, where a shadow that spans more than 180$\degr$ is attributed to two misaligned components \citep{Muro-Arena2020}. The location of the shadows and outer disk height place constraints on the inner disk orientation \citep{Min2017}. Evidence for disk tearing in a triple star system has been reported in GW Orionis \citep{Kraus2020}. Depending on the inner disk precession rate, these sources can exhibit variability. Multi-epoch observations of J1604 revealed variability of a narrow shadow \citep{Pinilla2018}. TW Hya is a unique case; observations of its shadows \citep{Debes2017} are also reported in the molecular line integrated intensity, but with an azimuthal offset \citep{Teague2022}. Furthermore, a multi-epoch observation study showed that the single moving shadow evolved into two separate ones, suggesting that there are two misaligned precessing components \citep{Debes2023}. 
    
    In molecular line observations, warps appear as a characteristic twist in the velocity maps \citep{Juhasz2017}. This distinct feature has been observed across various molecular tracers, including $^{12}$CO$(J=2-1)$ \citep{Perez2018}, $^{12}$CO$(J=3-2)$ \citep{Rosenfeld2014}, HCO$^{+}(J=3-2)$ \citep{Loomis2017}, HCO$^{+}(J=4-3)$ \citep{Rosenfeld2014}, and $^{12}$CO$(J=6-5)$ \citep{Casassus2015}, and it has been reported by multiple studies: TW Hya \citep{Rosenfeld2012}, HD142527 \citep{Casassus2013, Casassus2015, Garg2021}, HD100546 \citep{Walsh2017}, AA Tau \citep{Loomis2017}, MWC 758 \citep{Boehler2018}, J1604 \citep{Mayama2018}, HD143006 \citep{Perez2018}, HD 100453 AB \citep{vanderPlas2019}, GW Orionis \citep{Kraus2020, Jiaqing2020}, Elias 2-27 \citep{Paneque2021}, IRS48 \citep{vanderMarel2021}, CG Tau \citep{Wolfer2021}, and SY Cha \citep{Orihara2023}. Unfortunately, radial flows produce a similar signature in the velocity isophotes \citep{Rosenfeld2014}, and therefore it is not clear from molecular line observations which phenomenon is present or dominant within the disk. Distinguishing between these event configurations is a nontrivial task solely based on molecular line observations, and the complexity of interpreting these observational signatures stresses the need for complementary modeling approaches to disentangle the underlying physical processes. 

    High spatial and spectral resolution observations with the Atacama Large Millimeter/submillimeter Array (ALMA) have enabled mapping the kinematic structures of protoplanetary disks. Kinematic observations, commonly referred to as gas kinematics, have emerged as a valuable tool for studying the disk velocity structure and perturbations from a Keplerian velocity field. The widely used package Extracting Disk Dynamics \texttt{eddy} \citep{Teague2019_eddy} fits a Keplerian rotation model to the velocity maps under the assumption of axial symmetry. However, this approach encounters limitations in its assumptions when attempting to characterize disks with radial and azimuthal variations of the position angle, PA, and inclination, $i$, in the presence of a warp. An alternative package that is considered for these variations is \texttt{ConeRot} \citep{Casassus2022_ConeRot}. It fits concentric annuli allowing for independent PA and $i$ at each radius, although its implementation does not include a method to account for surface shadowing, which can lead to ambiguities, particularly in regions where the disk is self-obscured, compromising the accuracy of the inferred kinematical features. A more advanced 3D modeling method is \texttt{DiscMiner} \citep{Izquierdo2021}. This tool is specifically designed for fitting intensity channel maps, offering a more detailed analysis. However, the current implementation for the line profile kernel only allows the inclusion of Keplerian components, overlooking the influence of the warped structure on the observed line profiles.

    In the context of gas kinematics, the common assumption of an azimuthally symmetric disk during model fitting introduces significant limitations that lead to inaccurate velocity field inferences. \cite{Young2022} suggests that evidence of kinematic warp may have been missed due to the conventional analysis methods. To this end, we present an extension of \texttt{eddy}, enabling the fitting of a parametric warp. This add-on aims to address the limitations of assuming azimuthal symmetry, allowing for a more refined exploration of the kinematic structures within circumstellar disks. In this approach, we take into account self-shadowing effects in the generation of velocity centroids, overcoming the drawbacks of previous packages. To validate our tool, we construct three types of radiative transfer models: a warped disk, a disk with a radial flow, and a model combining a warped disk with a radial flow. We explore the residual patterns arising after fitting a warped disk to each radiative model, and discuss the particular observational features as a criteria to distinguish between cases.

    Our work is presented as follows: In Sect. \ref{sec:Methods} we explain the prescription used for warping a coordinate system, the disk model adopted, the setup for the radiative transfer code, how we create synthetic observations, and the extension to \texttt{eddy} to account for warp curve fitting. The velocity maps, as well as the resulting residuals are presented in Sect. \ref{sec:Results}. In Sect. \ref{sec:Discussion} we discuss the validity and limitations of the method, and in Sect. \ref{sec:Conclusions} we summarize and conclude.
    
\section{Methods} 
\label{sec:Methods}
        
    \subsection{Warp model and coordinates - Sky description}
    \label{sec:warp_coordinates}

        A disk can be modeled as a series of concentric annuli \citep{1974Rogstad, Abedi2014, Kamphuis2015}, where each annulus is characterized by an inclination ($i$) and a position angle (PA). Establishing a suitable reference frame is essential for developing a model of a warped disk. In this work, we explore two descriptions of warped disks.
        
        In a sky description \citep[e.g.,][]{Bohn2022}, we establish a coordinate system where the x-axis points north, the y-axis points east, and the z-axis extends toward the observer at infinity. A warped disk can be constructed based on a parametrization of the normal vector describing each annulus. We define $i(r) = a_{\mathrm{sky}}\left (r, i_{\mathrm{in}}, i_{\mathrm{out}}, a_{dr}, a_{r_0} \right )$ as a rotation with respect to the x-axis of the sky and PA$(r) = a_{\mathrm{sky}}\left (r, \mathrm{PA}_{\mathrm{in}}, \mathrm{PA}_{\mathrm{out}}, a_{dr}, a_{r_0} \right ) $ as a rotation around the z-axis, representing the observed major axis associated with each annulus. Here, $r$ denotes the midplane radial coordinate. We define a radial profile that smoothly transitions from $a_{\text{in}}$ to $a_{\text{out}}$ across a distance of $a_{\text{in}}$, given by:

        \begin{equation}
            a_{\mathrm{sky}}\left (r, \vec{p}_{\text{sky}} \right ) = a_{\mathrm{out}} - \frac{\left ( a_{\mathrm{out}} - a_{\mathrm{in}} \right )}{\left ( 1 + \exp{\left ( \frac{r - a_{r_0}}{0.1a_{\text{dr}}} \right )} \right )}, 
            \label{eq:logistic_sky}
        \end{equation}
        
        \noindent where $\vec{p}_{\text{sky}} = \left \{a_{\text{in}}, a_{\text{out}}, a_{dr}, a_{r_0}  \right \}$ is the parameter vector, $a_{\mathrm{in}}$ is a parameter that sets the inner disk PA and $i$, $a_{\mathrm{out}}$ sets the outer disk PA and $i$, $a_{r_0}$ specifies the inflection point of the curve, representing the midpoint of the transition, and $a_{dr}$ controls the width of the curve, adjusting the rate of transition.
        
        The coordinate transformation for these disks is described by the rotation matrices $R_z(\mathrm{PA}(r))$ and $R_x(i(r))$, accounting for the radial dependence of these angles.

        \begin{equation}
            \left ( {x}'_{w}, {y}'_{w}, {z}'_{w} \right )^{T} = R_z(\mathrm{PA}(r)) \cdot R_x(i(r))  \cdot \left ( x, y, z \right )^{T},
        \end{equation}

        \noindent where $\cdot$ denotes matrix multiplication. The terms $R_z(\mathrm{PA}(r))$ and $R_x(i(r))$ correspond to the rotation matrices around the z and x axes, respectively. The transformed Cartesian coordinates in the warped frame are denoted as ${x}'_{w}, {y}'_{w}, \text{and } {z}'_{w}$. The rotation matrices are defined by

        \begin{equation}
            R_z\left ( \theta \right ) =  \begin{bmatrix}
                                             \cos{\theta} & -\sin{\theta} & 0 \\ 
                                             \sin{\theta} & \ \ \cos{\theta} & 0\\ 
                                             0 & 0 & 1 
                                         \end{bmatrix}
        \end{equation}

        \begin{equation}
            R_x\left ( \theta \right ) =  \begin{bmatrix}
                                                 1 & 0 & 0\\ 
                                                 0 & \cos{\theta} & -\sin{\theta} \\ 
                                                 0 & \sin{\theta} & \ \ \cos{\theta} 
                                            \end{bmatrix}.
        \end{equation}

        \subsection{Warp model and coordinates - Disk description}
        
        Warps can also be described in relation to the outer disk, where the orientation of the inner disk is specified relative to the outer disk. We introduce the tilt angle $\beta(r) = a_{\mathrm{disk}}\left (r, \beta_{\mathrm{in}}, a_{dr}, a_{r_0} \right ) $, representing the relative inclination between the inner and outer disk midplanes, and the twist angle $\gamma(r) = a_{\mathrm{disk}}\left (r, \gamma_{\mathrm{in}}, a_{dr}, a_{r_0} \right )$, denoting the rotation relative to normal vector of the outer disk. With its associated parameter vector as $\vec{p}_{\text{disk}} = \left \{ \text{a}_{\text{in}}, a_{dr}, a_{r_0}  \right \}$. The function describing the radial profile in this case is given by
        
        \begin{equation}
            a_{\mathrm{disk}}\left (r, \vec{p}_{\text{disk}} \right ) = \frac{a_{\mathrm{in}}}{\left ( 1 + \exp{\left ( \frac{r-a_{r_0}}{0.1a_{\text{dr}}} \right )} \right )},
            \label{eq:logistic_disk}
        \end{equation}
        
        \noindent where $r$ is the radius of the unperturbed disk, $a_{\mathrm{in}}$ is a parameter that sets the inner disk tilt or twist. We note that the curves share the inflection point $a_{r_0}$ and the transition width $a_{dr}$.
        
        Two additional transformations, accounting for the outer disk PA and inclination, are applied. In this case, the warped coordinates are defined as follows:

        \begin{equation}
            \left ( {x}'_{w}, {y}'_{w}, {z}'_{w} \right )^{T} = R_z(\mathrm{PA}) \cdot R_x(\mathrm{inc}) \cdot R_z(\gamma(r)) \cdot R_x(\beta(r)) \cdot \left ( x, y, z \right )^{T}.
        \end{equation}
        
\begin{figure*}
        \centering
            \includegraphics[width=17cm]{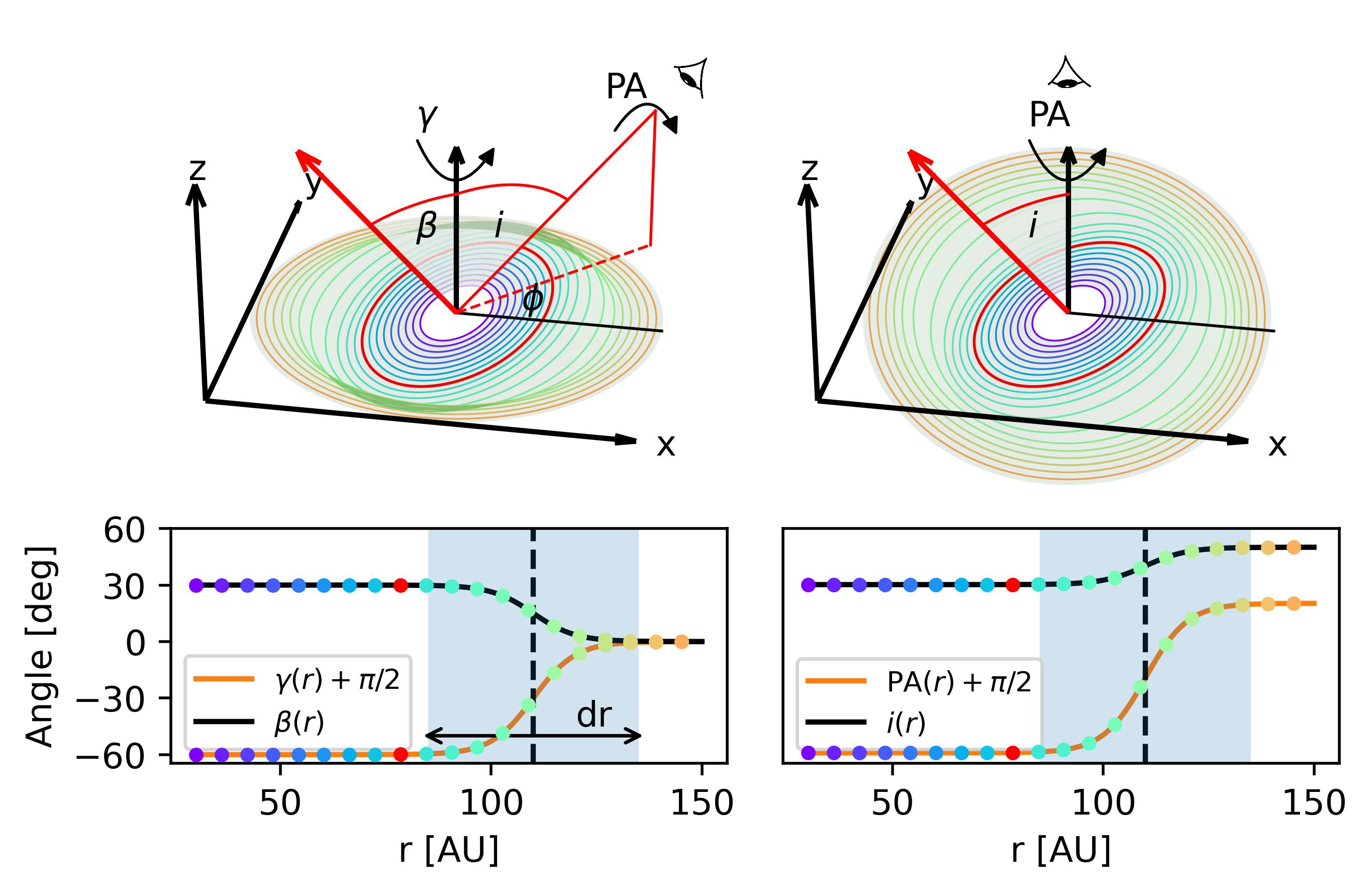}
            \caption{Geometrical representation of the warp models for each prescription. Left: Disk-frame representation of the warp. The inner disk orientation is characterized by the angle $\beta$, representing the inclination with respect to the outer disk normal vector, and the twist angle $\gamma$ denoting the rotation around the normal vector. The direction of the observer is defined through a rotation around the outer disk z-axis by $\phi$, and inclination $i$ away from it, the camera is rotated by PA around the vector pointing toward the observer. Right: Sky-frame representation of the warp. Each ring has an associated $i$ and PA. The inclination is defined with respect to the z-axis. The PA rotation is applied around the line-of-sight represented by the black arrow. On the bottom the respective profiles are plotted. The colored dots represent each ring position, the inflection point is demarcated by the black dashed line, and the rectangular region shows the transition width. We note that the angles shown are representative of the highlighted red ring, the red arrow is the normal vector to the surface defined by that ring.}
            \label{fig:warp_representation}
        \end{figure*}

        A visual representation of a warped disk model for each description, along with their respective profiles, is presented in Fig. \ref{fig:warp_representation}.

        When comparing these descriptions, it is important to note their intrinsic differences. The sky description establishes orientation relative to the observer direction, while the disk description represents the orientation of the inner disk in terms of the outer disk. Each description is characterized by distinct angles, therefore, these descriptions are distinct. However, they can be represented within the other framework. The selection of the appropriate framework depends on the specific modeling requirements.
        
    \subsection{Disk model}
    \label{sec:disk_model}

        To assess the efficacy of the proposed method, we apply it to simulated data that represent the ground truth our model aims to recover. In the following, we explain how we generate three distinct model types: a warped disk, an unperturbed disk with radial flow, and a warped disk with a radial flow. The vertical density structure is described by a Gaussian profile, defined as:

        \begin{equation}
            \rho\left ( {x}'_{u}, {y}'_{u}, {z}'_{u} \right ) = \frac{\Sigma\left ( {x}'_{u}, {y}'_{u} \right )}{H_{\mathrm{p}}\left ( {x}'_{u}, {y}'_{u} \right )\sqrt{2\pi}}\mathrm{exp}\left ( -\frac{z_{u}^{\prime 2}}{2H_{\mathrm{p}}\left ( {x}'_{u}, {y}'_{u} \right )^{2}} \right ),
        \end{equation}

        \noindent where $({x}'_u, {y}'_u, {z}'_u)$ denote the unwarped coordinates: $\Sigma\left ( {x}'_u, {y}'_u \right )$ is the surface density profile, and $H_{\mathrm{p}}\left ( {x}'_u, {y}'_u \right )$ represents the pressure scale height. We assume that these two distributions follow a power-law profile, which can be expressed as:

        \begin{equation}
            \Sigma \left ( {x}'_u, {y}'_u \right ) = \Sigma_0 \left ( \frac{R\left ( {x}'_u, {y}'_u \right )}{r_0} \right )^{-\gamma},
        \end{equation}

        \begin{equation}
            H_{\mathrm{p}}\left ( {x}'_u, {y}'_u \right ) = H_0 \left ( \frac{R\left ( {x}'_u, {y}'_u \right )}{r_0} \right )^{\zeta } R\left ( {x}'_u, {y}'_u \right ),
        \end{equation}
        
        \noindent where $r_0$ is the reference radius set to $\SI{1}{au}$, while $\Sigma_0$ and $H_0$ represent the surface density and the aspect ratio at that radius, respectively. The gas surface density exponent is denoted by $\gamma = 1$, and the flaring index is $\zeta = 0.25$. We have not incorporated an exponential taper in the surface density profile because it does not affect the regions we are interested in.

        All models assume a Keplerian velocity field. For models that include a radial flow, we follow a similar approach as outlined in \citep{Rosenfeld2014}, where an inward radial velocity component is introduced as a fraction $\chi\left ( r \right )$ of the local Keplerian velocity. A value $\chi = 0$ corresponds to a perfect circular Keplerian velocity, while $\chi = 1$ indicates adding a radial inward velocity with a magnitude equal to the local Keplerian speed. In this case, the velocity field is:

        \begin{equation}
            v_x = -v_{\mathrm{K}}\sin{\phi} - \chi\left ( r \right )v_{\mathrm{K}}\cos{\phi}
        \end{equation}

        \begin{equation}
            v_y = v_{\mathrm{K}}\cos{\phi} - \chi\left ( r \right )v_{\mathrm{K}}\sin{\phi},
        \end{equation}

        \noindent where $\phi$ is the azimuthal angle in the disk plane. We adopt the same functional form as given in Eq. \ref{eq:logistic_disk} for the function $\chi \left ( r \right)$. It is expressed as $\chi \left ( r \right) = a_{\mathrm{disk}}\left ( r, \chi_{0}, \chi_{r_0}, \chi_{dr} \right )$.
        
        The function describes a smooth curve that decreases monotonically from the scaling factor $\chi_{0}$ to 0, radially outward. The net effect is that it introduces a radial velocity gradually, generating an incremental twist in the isovelocity curves. We have fixed the scaling factor $\chi_{0} = 1$, following the procedure by \cite{Rosenfeld2014}. However, the scaling factor can take any value within the range of 0 (indicating an absence of radial flows) to $\sqrt{2}$ (radial infall at the local free-fall velocity). We emphasize that the parameters $\chi_{r_0}$ and $\chi_{dr}$ are independent of those defining the warp. A radial flow has a similar visual appearance in the velocity maps to warps.

        \subsection{Radiative transfer simulations}

        We use the 3D radiative transfer simulation code \texttt{RADMC3D}\footnote{ \url{http://www.ita.uni-heidelberg.de/dullemond/software/radmc-3d/}} \citep{Dullemond2012} to model the molecular line emission from our prescribed disk configurations. A central star with a temperature of 5780K, a radius of 1 R$_{\odot}$, mass of 1 M$_{\odot}$ is placed at the grid center, while we set the system at a distance of 140~pc. We used the \texttt{mctherm} task to compute the dust temperatures in the disk and assumed them equal to the gas temperature. Line radiative transfer calculations were conducted under the assumption of local thermodynamic equilibrium (LTE). The molecular data are obtained from the LAMDA database \citep{Schoier2005}. We included microscopic turbulence, which approximately follows $\sqrt{\alpha_{\text{SS}}}c_{s}$, where $c_{s}$ is the sound speed and $\alpha_{\text{SS}}$ is the viscosity parameter. We assumed $\alpha_{\text{SS}} = 10^{-3}$. For the $^{12}$CO molecular abundance, we assumed the typical 10$^{-4}$ fraction relative to H$_{2}$ \citep{Young1991, Lacy1994}. In regions where $T < \SI{20}{\kelvin}$, we set the CO abundance to 0 to mimic freeze-out. We did not include self-shielding of CO.

        The models were implemented using spherical coordinates, where we initialize $N_r$ = 100, $N_\theta$ = 180, and $N_\phi$ = 180 grid points in the radial, azimuthal, and polar directions, respectively. The radial dimension of the mesh spans from $\SI{0.5}{au}$ to $\SI{100}{au}$, using a logarithmic distribution. The azimuthal and polar directions are distributed linearly in the ranges $\left [ 0, 2\pi \right ]$ and $\left [ 0, \pi \right ]$, respectively. An additional refinement is included in the models that present a warped disk, adding 50 cells in the radial direction. These cells are linearly distributed within the range covering $\left [a_{r_0} - a_{dr}, a_{r_0} + a_{dr} \right ]$, thus the radial grid accounts for the sudden change in orientation from the inner to the outer disk. In case of a disk exhibiting only a radial flow, these cells are distributed within the range $\left [\chi_{r_0} - \chi_{dr}, \chi_{r_0} + \chi_{dr} \right ]$.

        We computed the channel maps for the $^{12}$CO$(J = 2-1)$ line transition centered at 230.538 GHz. The velocity range considered was $\pm$15 km s$^{-1}$, with a channel spacing of 40 m s$^{-1}$, comparable to the maximum velocity resolution achieved by ALMA in Band 6.

        To assess the observability of these structures, we generated synthetic observations from our models. We employed a two-step post-processing procedure to distinguish between the impacts of spatial resolution and thermal noise. In the first step, a circular Gaussian beam was convolved with each data cube. In the second step, a Gaussian noise image was created, convolved with the same beam, rescaled to the characteristic sensitivity of exoALMA, and finally added to the convolved images. We used a $0.03\arcsec$ FWHM Gaussian beam for the test cases. This represents an optimistic scenario for the observable kinematic signatures achievable with ALMA. For the models shown in Sect. \ref{sec:Results}, we incorporated noise with a root mean square (RMS) of $\SI{3.0}{\kelvin}$ over $\SI{150}{\metre\per\second}$, similar to the exoALMA setup. Our analysis omits consideration of potential image artifacts arising from sparse uv coverage.

    \subsection{Warp extension for \texttt{eddy}}

        The \texttt{eddy}\footnote{\url{https://github.com/richteague/eddy/}} package is a tool designed for extracting kinematical information from spatially and spectrally resolved line data by fitting a Keplerian rotation pattern \citep{Teague2019_eddy}. This package performs a fitting of the first moment map of an optically thick line by modeling the projected rotation velocity $v_{\phi,\mathrm{proj}}(r,\phi) = v_{\phi}(r)\cos{\phi}\sin{i}$ to recover the geometrical parameters of the disk and identify localized deviations in the velocity structure. This procedure assumes an axisymmetric disk structure. Hence, it does not account for a radial variation of the geometrical parameters. The package has been largely employed to determine the geometry of disks, as well as to identify localized deviations from the Keplerian field \citep[see e.g.,][]{Galloway-Sprietsma2023, Ribas2023, Garg2022}.
        
        Our extension introduces the capability to incorporate both parametric warp prescriptions and a radial flow\footnote{Available to the public at \url{https://github.com/andres-zuleta/eddy/tree/warp_rf}}, as described in Sect. \ref{sec:warp_coordinates}. To address the shadowing problem that arises from projecting a 3D surface to 2D, we implemented an algorithm based on ray tracing. The algorithm takes two matrices with image dimensions n$_x$ and n$_y$, providing the position of these pixels, a set of 2D meshes containing the $({x}'$, ${y}'$, ${z}')$ positions with corresponding warped surface values, and a 3D matrix containing the quantities for interpolation at pixel positions as an input. To define a surface, a triangulation is applied to the set of points defining the disk. This triangulation defines the vertices of the simplices as adjacent grid points. Then, for each image pixel, a ray is cast from infinity toward -z, to search for an intersection with the triangles, and determine the closest intersected triangle to the image plane. For each selected surface, a barycentric interpolation is employed to approximate the object value at the image pixel position. This process enables the recovery of projected and interpolated values into a 2D image view, addressing the challenges posed by the projection of 3D structures onto 2D observational data. This represents a significant improvement over the \texttt{shadowed} approach already implemented in \texttt{eddy}, providing a more accurate and robust method for deprojecting pixel values describing complex 3D surfaces from observational data.
        
\section{Results}
\label{sec:Results}

    This work extends the current capabilities of characterizing astrophysical disks using gas kinematics by incorporating warped geometries into \texttt{eddy}. For the upcoming results, we generate models using the sky prescription, with standard parameters set as $\text{PA}_{\text{out}} = 0\degr$ and $i_{\text{out}} = 25\degr$, defining a prograde disk. We first introduce an inclination-only warp model, where only the inclination changes with radius while the position angle remains fixed. We set $i_{\text{in}} = 35\degr$, allowing us to focus solely on the effects of inclination variations.
    
    For warped and twisted disks, we adopt $\text{PA}_{\text{in}} = -45\degr$, $i_{\text{in}} = 35\degr$. For warped disks with radial flows, we use generic values of $\text{PA}_{\text{in}} = 10\degr$, $i_{\text{in}} = 40\degr$.
    
    All models share common warp parameters, with a warp position set at $a_{r_0} = \SI{40.0}{au}$ and a transition length of $a_{dr} = \SI{20.0}{au}$. For disks with radial flow, we chose $\chi_{r_0} = \SI{40.0}{au}$ and $\chi_{dr} = \SI{20.0}{au}$. These values ensure similarity in the morphology of the velocity maps.
    
    \cite{Rosenfeld2014} demonstrated that employing high angular and spectral resolution observations of molecular lines enables the detection of inflows, which manifest as twisted isophotes proximal to the systemic velocity. However, it was noted that other mechanisms, such as warping, can produce the same pattern. To illustrate the inherent degeneracy in the signals between radial flows and warps, we show the channel maps of the three distinct types of models described in Sect. \ref{sec:disk_model} in Fig. \ref{fig:channel_warps}. From now on, we refer to these models as Model W (warped disk model), Model F (disk with radial flows), and Model WF (warped disk model with radial flows). All models exhibit a similar overall physical shape in the first moment map, featuring a distinctive twisted emission pattern typical of sources with non-planar or radial motions.

    \subsection{Simple models}
    \label{sec:simple_models}

\begin{figure*}
            \centering
            \includegraphics[width=17cm]{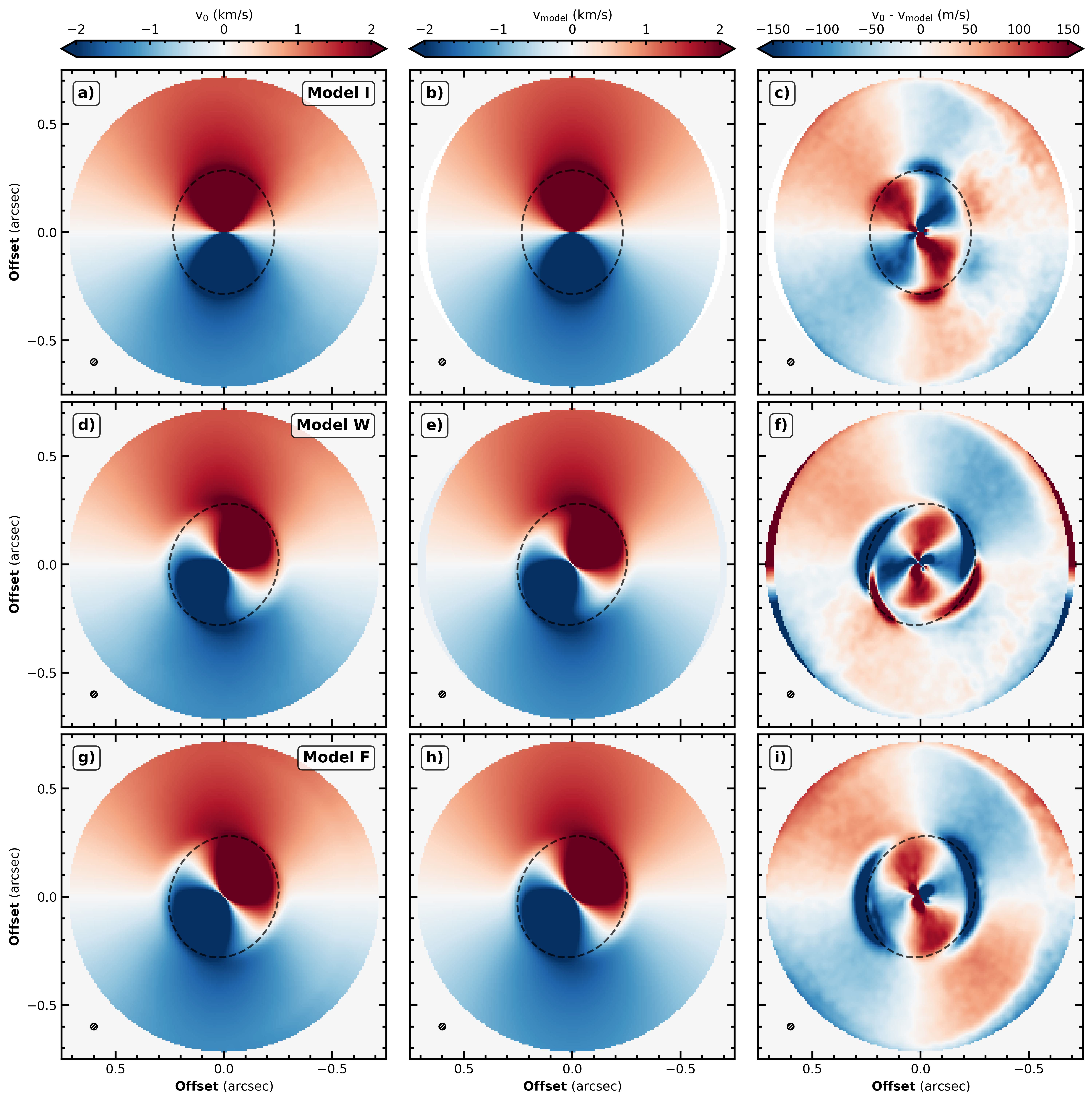}
            \caption{Gallery of $v_{0}$ velocity maps for Model I, W, and F, indicated in the top right corner. The left column displays the synthetic velocity maps, the middle column shows the respective best-fit, and the right column illustrates the residuals after subtracting a best-fit model. The beam size is shown in the lower left corner of each panel, and the black dashed line marks the radius at which the transition occurs.}
            \label{fig:models_WF}
        \end{figure*}

        To generate velocity maps, we utilize the \texttt{bettermoments}\footnote{\url{https://github.com/richteague/bettermoments}} \citep{Teague2018} package, employing a Gaussian fit. To validate whether the inclination-only warp, warp model, and radial flow implemented in \texttt{eddy} accurately recover the input warp and radial flow, we conduct a fit to each synthetic model velocity maps with its corresponding mode. In Fig. \ref{fig:models_WF}, we present the velocity map of the synthetic observations, alongside their respective best fits. The free parameters considered are: $\Theta_{I} = \left \{i_{\text{in}}, \text{PA}_{\text{in}}, a_{dr}, a_{r_0}  \right \}$ for the inclination-only warp, $\Theta_{W} = \left \{ \text{PA}_{\text{in}}, i_{\text{in}}, a_{dr}, a_{r_0}  \right \}$ for the warp, and $\Theta_{F} = \left \{\text{PA}_{\text{out}}, i_{\text{out}}, \chi_{dr}, \chi_{r_0}  \right \}$ for the radial flow. We restrict the free parameters to those describing the warp or radial flow, respectively, for computational efficiency, however we ensured that this decision does not affect the posterior distribution. We then sampled the posterior using an affine-invariant Markov Chain Monte Carlo (MCMC) \citep{Goodman2010} fitting routine implemented with \texttt{emcee} \citep{Foreman-Mackey2013}. We adopted 32 walkers, 1,000 burn-in steps, and 1,000 steps. The total number of steps ensures integrated autocorrelation time values between 20 and 90 for the chain of each parameter. Visual inspection of the chains confirms walker convergence. We then take $N$ random draws from the sample, and average them to generate a model rotation map, denoted as $v_{\text{model}}$. Across all synthetic models, we observe a twisted pattern in the velocity map with a smooth transition to the outer disk.

        For Model I, the posterior distribution reveals an inner inclination of $i_{\text{in}} = 34^{\circ}.32_{-7.7\text{E}-04}^{+7.7\text{E}-04}$ and $\text{PA}_{\text{in}} = 0^{\circ}.02_{-1.1\text{E}-03}^{+9.8\text{E}-04}$. The inflection point is found at $a_{r_0} = 0.29\arcsec\ _{-1.1\text{E}-05}^{+1.2\text{E}-05}$ (equivalent to \SI{40.96}{au}), and the transition width is $a_{dr} = 0.20\arcsec\ _{-9.5\text{E}-05}^{+9.4\text{E}-05}$ (\SI{27.67}{au}). These results are listed in Table \ref{table:1}. In our posterior distributions, we encounter uncertainties that are notably small, often on the order of $10^{-3}$ or smaller. This issue is discussed in Sect. \ref{sec:Discussion}.
    
        For Model W, we obtain a posterior distribution spanning an inner position angle $\text{PA}_{\text{in}} = -44^{\circ}.75_{-1.3\text{E}-03}^{+1.3\text{E}-03}$, inner inclination $i_{\text{in}} = 33^{\circ}.98_{-6.5\text{E}-04}^{+6.6\text{E}-04}$, inflection point $a_{r_0} = 0.28\arcsec\ _{-4.4\text{E}-06}^{+4.1\text{E}-06}$ (equivalent to \SI{38.87}{au}), and transition width $a_{dr} = 0.18\arcsec\ _{-3.0\text{E}-05}^{+2.9\text{E}-05}$ (\SI{25.22}{au}).
    
        For the Model F, we obtain $\text{PA}_{\text{out}} = -0^{\circ}.06_{-5.8\text{E}-04}^{+5.8\text{E}-04}$, outer inclination $i_{\text{out}} = 24^{\circ}.72_{-2.5\text{E}-04}^{+2.6\text{E}-04}$, radial flow inflection point $\chi_{r_0} = 0.28\arcsec\ _{-4.4\text{E}-06}^{+4.2\text{E}-06}$ (equivalent to \SI{39.74}{au}), and transition width $\chi_{dr} = 0.19\arcsec\ _{-2.9\text{E}-05}^{+3.0\text{E}-05}$ (\SI{26.42}{au}). The findings are summarized in Table \ref{table:1}.
        
        The residuals reveal two main features (Fig. \ref{fig:models_WF}f, \ref{fig:models_WF}i): a central multipole attributed to high-velocity gas components outside the synthetic cube velocity range, and a double arc feature displaying sub- and super-Keplerian rotation at the transition length that emerges since in the radiative transfer model the emission comes from a emitting volume, while the sampler approximates this region as a surface. Minor scale deviations are attributed to injected thermal noise and the disk surface.
        
        We repeated the analysis using the same synthetic data, but convolved with a 0.1\arcsec FWHM beam, representing the angular resolution achieved by exoALMA. For Model I, we found an inner inclination $i_{\text{in}} = 32^{\circ}.66_{-5.1\text{E}-04}^{+5.3\text{E}-04}$, an inner position angle $\text{PA}_{\text{in}} = 0^{\circ}.01_{-6.9\text{E}-04}^{+7.0\text{E}-04}$, an inflection point $a_{r_0} = 0.29\arcsec\ _{-7.4\text{E}-06}^{+7.6\text{E}-06}$ (equivalent to \SI{41.07}{au}), and a transition width $a_{dr} = 0.22\arcsec\ _{-5.6\text{E}-05}^{+5.4\text{E}-05}$ (\SI{30.75}{au}). For the Model W, we found $\text{PA}_{\text{in}} = -43^{\circ}.19_{-1.2\text{E}-03}^{+1.3\text{E}-03}$, an inner inclination $i_{\text{in}} = 31^{\circ}.82_{-4.5\text{E}-04}^{+4.2\text{E}-04}$, an inflection point $a_{r_0} = 0.28\arcsec\ _{-3.5\text{E}-06}^{+3.5\text{E}-06}$ (\SI{38.72}{au}), and transition width $a_{dr} = 0.23\arcsec\ _{-2.1\text{E}-05}^{+2.2\text{E}-05}$ (\SI{32.69}{au}). For Model F, we found $\text{PA}_{\text{out}} = -0^{\circ}.01_{-2.6\text{E}-04}^{+2.7\text{E}-04}$, an outer inclination $i_{\text{out}} = 24^{\circ}.55_{-1.0\text{E}-04}^{+1.0\text{E}-04}$, a radial flow inflection point $\chi_{r_0} = 0.28\arcsec\ _{-3.0\text{E}-06}^{+2.8\text{E}-06}$ (equivalent to \SI{38.93}{au}), and transition width $\chi_{dr} = 0.27\arcsec\ _{-2.1\text{E}-05}^{+2.0\text{E}-05}$ (\SI{37.68}{au}). The recovered posteriors exhibited greater deviation from the input values compared to the 0.03\arcsec FWHM case from the input, with parameters $a_{r_0}$ and $\chi_{dr}$ being overestimated by approximately 25\% to 50\% (see Table \ref{table:1}). Upon examining the residuals (see Fig. \ref{fig:models_WF_0.1FWHM}), we observe a significant deviation from the background noise, indicating that our ability to characterize the kinematics of the system is limited by angular resolution.

        \bgroup
        \def\arraystretch{1.5}
        \begin{table}
        \caption{Posterior distribution of best-fit models.}              
        \label{table:1}      
        \centering                                      
        \begin{tabular}{l r r}          
        \hline\hline                        
        Model I Parameter             & 0.03\arcsec\ FWHM                                    & 0.1\arcsec\ FWHM \\    
        \hline                                   
        PA$_{\text{in}}$ \hfill (deg) & $0^{\circ}.02_{-1.1\text{E}-03}^{+9.8\text{E}-04}$   & $ 0^{\circ}.01_{-6.9\text{E}-04}^{+7.0\text{E}-04} $\\
        $i_{\text{in}}$ \hfill (deg)  & $ 34^{\circ}.32_{-7.7\text{E}-04}^{+7.7\text{E}-04}$ & $ 32^{\circ}.66_{-5.1\text{E}-04}^{+5.3\text{E}-04}$ \\
        $a_{dr}$ \hfill (arcsec)      & $ 0.20\arcsec\ _{-9.5\text{E}-05}^{+9.4\text{E}-05}$ & $ 0.22\arcsec\ _{-5.6\text{E}-05}^{+5.4\text{E}-05}$  \\
        $a_{r_0}$ \hfill (arcsec)     & $ 0.29\arcsec\ _{-1.1\text{E}-05}^{+1.2\text{E}-05}$ & $ 0.29\arcsec\ _{-7.4\text{E}-06}^{+7.6\text{E}-06}$ \\
        \hline\hline                        
        Model W Parameter             & 0.03\arcsec\ FWHM                                    & 0.1\arcsec\ FWHM \\    
        \hline                                   
        PA$_{\text{in}}$ \hfill (deg) & $-44^{\circ}.75_{-1.3\text{E}-03}^{+1.3\text{E}-03}$ & $-43^{\circ}.19_{-1.2\text{E}-03}^{+1.3\text{E}-03}$\\
        $i_{\text{in}}$ \hfill (deg)  & $ 33^{\circ}.98_{-6.5\text{E}-04}^{+6.6\text{E}-04}$ & $ 31^{\circ}.82_{-4.5\text{E}-04}^{+4.2\text{E}-04}$ \\
        $a_{dr}$ \hfill (arcsec)      & $ 0.18\arcsec\ _{-3.0\text{E}-05}^{+2.9\text{E}-05}$ & $ 0.23\arcsec\ _{-2.1\text{E}-05}^{+2.2\text{E}-05}$  \\
        $a_{r_0}$ \hfill (arcsec)     & $ 0.28\arcsec\ _{-4.4\text{E}-06}^{+4.1\text{E}-06}$ & $ 0.28\arcsec\ _{-3.5\text{E}-06}^{+3.5\text{E}-06}$ \\
        \hline\hline
        Model F Parameter & & \\
        \hline
        PA$_{\text{out}}$ \hfill (deg) & $0^{\circ}.06_{-5.8\text{E}-04}^{+5.8\text{E}-04}$ & $-0^{\circ}.01_{-2.6\text{E}-04}^{+2.7\text{E}-04}$\\
        $i_{\text{out}}$ \hfill (deg)  & $ 24^{\circ}.72_{-2.5\text{E}-04}^{+2.6\text{E}-04}$ & $ 24^{\circ}.55_{-1.0\text{E}-04}^{+1.0\text{E}-04}$ \\
        $\chi_{dr}$ \hfill (arcsec)    & $ 0.19\arcsec\ _{-2.9\text{E}-05}^{+3.0\text{E}-05}$ & $ 0.27\arcsec\ _{-2.1\text{E}-05}^{+2.0\text{E}-05}$  \\
        $\chi_{r_0}$ \hfill (arcsec)   & $ 0.28\arcsec\ _{-4.4\text{E}-06}^{+4.2\text{E}-06}$ & $ 0.28\arcsec\ _{-3.0\text{E}-06}^{+2.8\text{E}-06}$ \\
        \hline
        \end{tabular}
        \tablefoot{
        Uncertainties correspond to the range between the 16th and 84th percentiles around the median value.}
        \end{table}
        \egroup

    \subsection{Warp and radial flow differentiation}

        Both a warp and radial flow can coexist in protoplanetary disks, complicating modeling efforts further. Previous attempts to fit a warp-only disk to twisted kinematical data have not fully explained the observed kinematical features in some systems \citep{Walsh2017, Loomis2017, Riviere-Marichalar2019}. To address these limitations, we adopt a compound warp and radial flow model in our \texttt{eddy} analysis. We assess the performance of \texttt{eddy} in recovering inputs from the combined warp and flow model (Model WF). The free parameters considered for this section are $\Theta_{WF} = \left \{ \text{PA}_{\text{in}}, i_{\text{in}}, a_{dr}, a_{r_0}, \chi_{dr}, \chi_{r_0} \right \}$.

\begin{figure*}
            \centering
            \includegraphics[width=17cm]{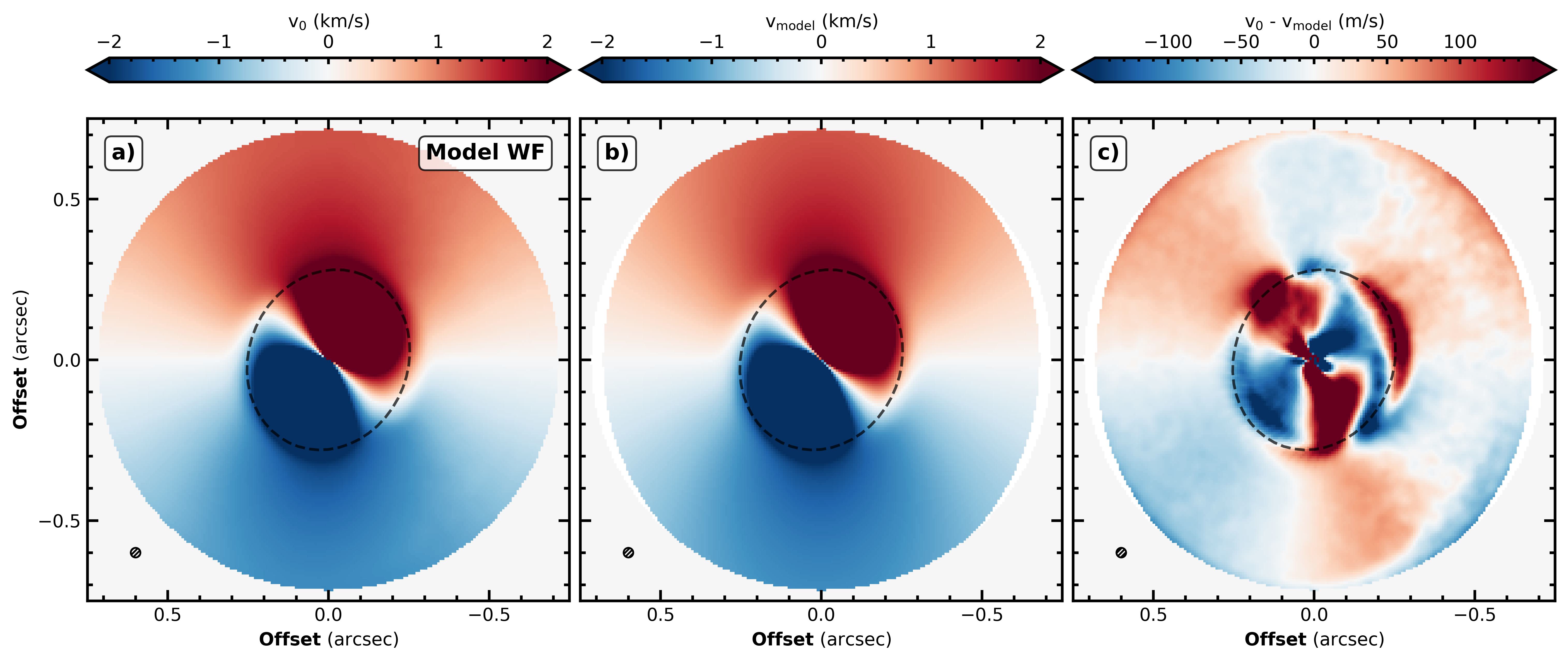}
            \caption{Velocity map $v_{0}$ for Model WF, along with its corresponding best-fit and residual maps. In each panel the beam size is indicated in the lower left corner. The black dashed line represents the radius at which the transition occurs.}
            \label{fig:models_warp_and_radialFlow}
        \end{figure*}

        Fitting the compound model in \texttt{eddy} to Model WF, we found $\text{PA}_{\text{in}} = 10^{\circ}.08_{-1.2\text{E}-03}^{+1.2\text{E}-03}$, an inner inclination $i_{\text{in}} = 38^{\circ}.89_{-9.5\text{E}-04}^{+1.0\text{E}-03}$, a transition width $a_{dr} = 0.18\arcsec\ _{-6.1\text{E}-05}^{+6.4\text{E}-05}$ (equivalent to \SI{25.19}{au}), an inflection point $a_{r_0} = 0.28\arcsec\ _{-1.2\text{E}-03}^{+1.2\text{E}-03}$ (\SI{39.71}{au}), a radial flow transition width $\chi_{r_0} = 0.29\arcsec\ _{-4.0\text{E}-06}^{+3.6\text{E}-06}$ (equivalent to \SI{40.83}{au}), and radial flow inflection point $\chi_{dr} = 0.19\arcsec\ _{-3.0\text{E}-05}^{+2.9\text{E}-05}$ (\SI{27.09}{au}). In Fig. \ref{fig:models_warp_and_radialFlow} we present a summary of these results. A similar multipole appears in the residuals (Fig. \ref{fig:models_warp_and_radialFlow}c). We observe that the residuals in the inner disk exceed the noise levels, with uncorrelated signals exhibiting non-informative kinematics.

        Additionally, we evaluate the results when fitting this combined model to a warp-only configuration (Model W) and radial flow-only (Model F). This general approach converges to the specific case of having solely a warp (i.e., $\chi_{r_0} \ll 0$) in Model W or a radial flow (i.e., $i_{\text{in}} \approx i_{\text{out}}$, and $\text{PA}_{\text{in}} \approx \text{PA}_{\text{out}}$) in Model F. The sensitivity of the routine highlights its capability to discern between different disk configurations.

\begin{figure}
            \resizebox{\hsize}{!}{\includegraphics{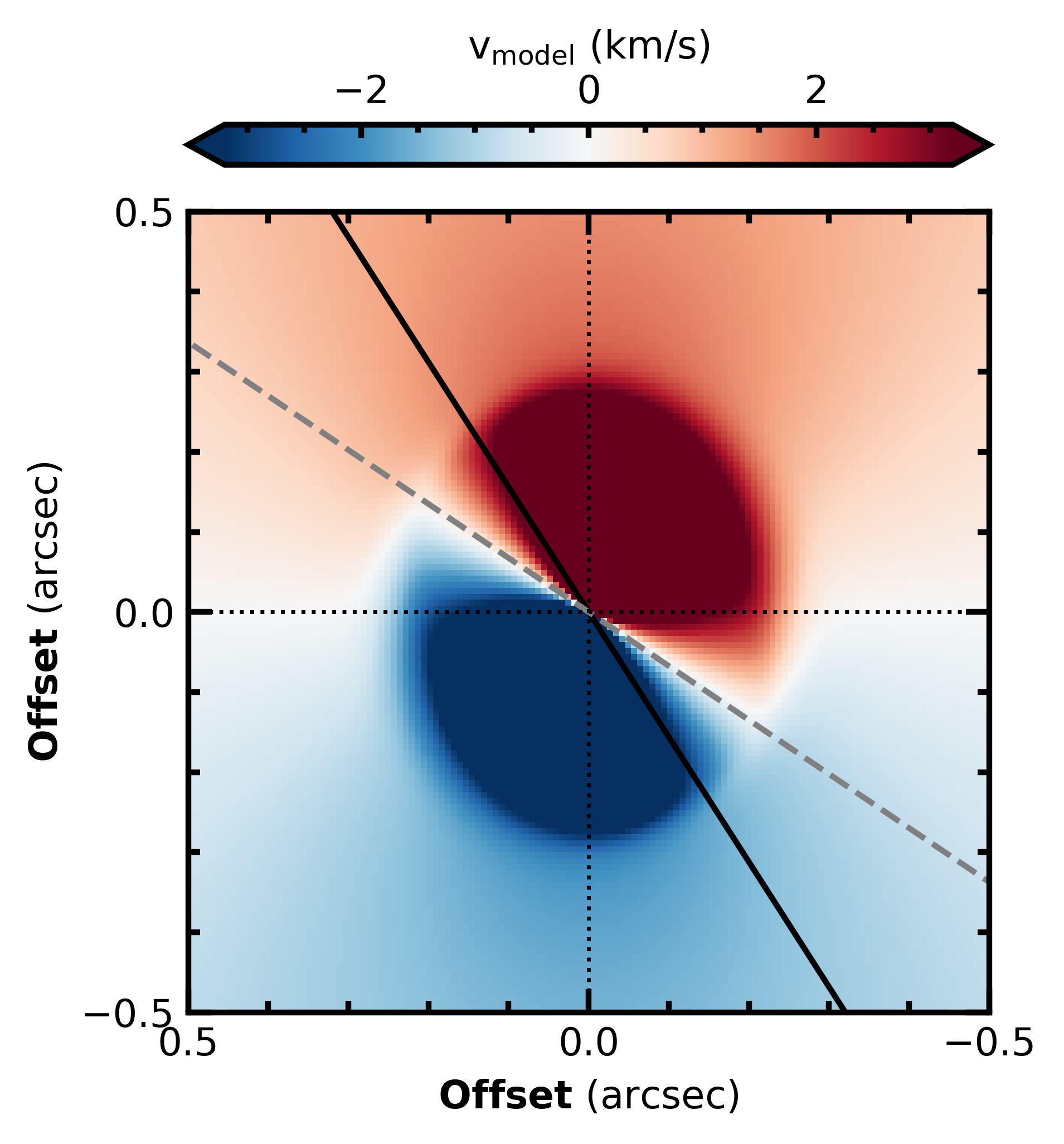}}
            \caption{Model velocity map $v_{0}$ for a disk with a radial flow and a warp. The solid black line represents the minor axis of a disk with a radial flow with the local free-fall velocity, while the dashed gray dashed line represents the actual inner minor axis.}
            \label{fig:maximum_RF}
        \end{figure}

        Another method to test for the presence of a warp in a system is to calculate the angle that defines the disk's minor axis, considering only a radial flow. This angle, denoted as ${\theta}'_{\text{sys}} = \arctan\left ( \chi^{-1}\cos{i} \right )$, is derived from Eq. 7 in \cite{Rosenfeld2014}\footnote{We note that in this derivation, the minor axis points north, rather than the east, as defined in our work.}. The equation shows a dependency on both the radial velocity and the inclination angle $i$. When gas moves at the local free-fall velocity, the parameters $\chi$ sets a maximum angle at which a radial flow can twist the disk, specifically when $\chi = \sqrt{2}$. Any angle that defines the minor axis and exceeds this threshold indicates the presence of a warp (see Fig. \ref{fig:maximum_RF}).

    \subsection{Diagnostics}
    \label{subsec:diagnostics}

        Distinguishing between the kinematic signatures of a warp, radial flow, or both is complex due to their similar imprints in the velocity maps. Without prior information about the system, misinterpreting the residuals from a velocity map modeling procedure is possible. To overcome this visual ambiguity, we search for characteristic signatures in the residuals when fitting a warp or radial flow to an intrinsically different model. By examining these residuals, we aim to identify unique patterns or deviations that indicate the underlying system kinematics. This approach aids in interpreting observed residual features as indicative of a warp, radial flow, or both.

\begin{figure}
            \resizebox{\hsize}{!}{\includegraphics{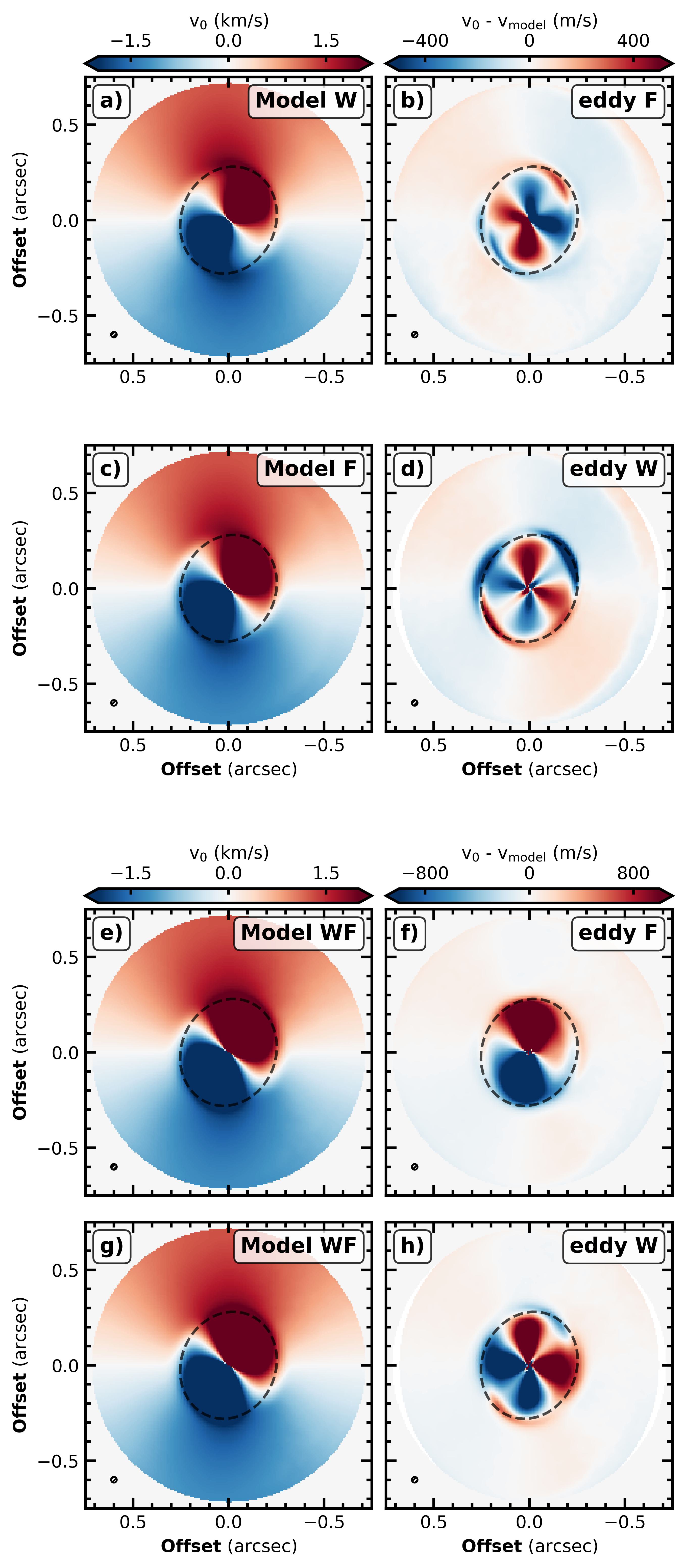}}
            \caption{Displaying $v_{0}$ galleries for various model types. The left column exhibits synthetic $v_{0}$ maps, with model types labeled in the top right corner. The right column displays residuals after subtraction of the best-fit model, also labeled in the top right corner. Beam size is denoted in the lower left corner of each peach panel. The black dashed line represents the radius at which the transition occurs.}
            \label{fig:models_diagnostics}
        \end{figure}

        A summary of the $v_{0}$ maps of the synthetic models is shown in the left column of Fig. \ref{fig:models_diagnostics}. The right column displays the residual after subtracting a best-fit \texttt{eddy} model, with the type of \texttt{eddy} model indicated in the top right corner. For \texttt{eddy} modes indicating a warp, we consider the following free parameters: $\Theta_{W} = \left \{ \text{PA}_{\text{in}}, i_{\text{in}}, a_{dr}, a_{r_0} \right \}$. For the mode indicating a radial flow, we considered: $\Theta_{F} = \left \{ \chi_{dr}, \chi_{r_0} \right \}$. 

        In Fig. \ref{fig:models_diagnostics}b, we observe the residuals after fitting a radial flow to a warped model. A strong quadrupole pattern is evident in the region defining inner disk. This systematic residual has a negative value towards inner disk major axis defining the red-shifted gas and a positive value towards the blue shifted part. Similarly, when fitting a warp to a model with a radial flow (Fig. \ref{fig:models_diagnostics}d), we identify a quadrupole with inverted values to those seen in Fig \ref{fig:models_diagnostics}b.

        The kinematics traced by Model WF are shown in Fig. \ref{fig:models_diagnostics}e and Fig. \ref{fig:models_diagnostics}g. In Fig. \ref{fig:models_diagnostics}f, a large perturbation defining a dipole is observed, matching the inner major axis. If we fit a warp (Fig. \ref{fig:models_diagnostics}h), a quadrupole is observed with a similar alignment to what is seen in Fig. \ref{fig:models_diagnostics}d. 

        Alongside these large deviations, two arcs at the boundary marking the transition from the inner to the outer region are identified. All residual patterns seem to be aligned with the inner disk major axis. The amplitude of the residual is significantly higher than the defined noise values.
        
        These features consistently emerge across models with different geometries within the same model type. Despite the convergence of \texttt{eddy}, the posterior distributions characterizing the geometrical parameters are not consistent with the synthetic data, as the model describes another system configuration.

\section{Discussion}
\label{sec:Discussion}

    Overall, a comprehensive analysis of the gas kinematics of a disk suspected of having warps, radial flows, or both, must be done adequately. The extension of the fitting routine enables the recovery of geometrical parameters only when data with sufficient angular resolution is available. If any of the signatures identified in Sect. \ref{subsec:diagnostics} arises, the fitted geometrical parameters should not be used, as they not align with the actual disk parameters. In such cases, the residuals should only be considered as an indicative of the kinematics present in the disk.
    
    The parameters obtained from our fitting procedure are consistent with the inputs of the synthetic radiative transfer models. However, two parameters, $a_{dr}$ and $\chi_{dr}$, exhibit a posterior distribution that deviates approximately 50\% from the input model. This discrepancy likely arises because our model assumes that the emission is coming from an emitting surface rather than a realistic emitting volume, leading to inaccuracies in the region defining the transition from the inner to the outer disk, where the double arc residual feature is present. Nevertheless, this offset does not significantly alter the results, and the  model remains valid for locations outside this region. The small uncertainties in the posterior distribution reflect the model underestimation of the data complexity, resulting in overly confident parameter estimates and consequently small uncertainties.

    Foreground absorption is another source of error in model fitting. This phenomenon is common in nearby molecular clouds \citep{vanKempen2009} and can significantly impact the analysis of the source. The presence of these foreground clouds distorts line profiles \citep{vanderMarel2013} and appears as reduced emission in the intensity maps. Such contamination can lead to inaccurate velocity maps. A documented example of this in a warped system is HD 142527 \citep{Casassus2015b}.
    
    Our ray-casting tool facilitates the deprojection of any type of surface. In this work, we focus on warped disks, characterized by misaligned inner and outer regions. By adjusting the inclination and position angle profiles to represent the desired geometry, the tool can be adapted to investigate the gas kinematics of more exotic systems. This could include torn disks, such as GW Ori \citep{Kraus2020}, or systems with multiple misalignments, such as HD 139614 \citep{Muro-Arena2020}.
    
    To evaluate the validity of the proposed method, we examined the effects of noise, spatial resolution, and inclination. In Fig.\ref{fig:models_highNoise}, we display the corresponding residuals after incorporating a noise level with an RMS of $\SI{3.5}{\kelvin}$ over $\SI{350}{\metre\per\second}$, similar to the observational setup of the Molecules with ALMA at Planet-forming Scales (MAPS) large program  \cite{2021_MAPS_I_in_press}. The results indicate that the contribution of the inclination $i$ to the amplitude of projected velocity surpasses the effects of thermal noise, suggesting that the validity of the method is not limited by noise constraints.
        
    In protoplanetary disks, warps have been reported at radii smaller than $\SI{20}{au}$, posing challenges for current observational capabilities due to the associated angular sizes. In Fig. \ref{fig:models_WF_0.1FWHM}, we show the corresponding residuals after subtracting the compatible best-fit model from the Model W and Model F data convolved with 0.1\arcsec\ FWHM beam. The posterior distributions recovered values comparable to those obtained in Sect. \ref{sec:simple_models}. However, the residuals are too large to confidently categorize the model type, leading to the potential misinterpretation of the features  as fitting the wrong model type to the data (see right column of Fig. \ref{fig:models_diagnostics}). We found that for an angular resolution finer than 0.04\arcsec, this effect is minimized, while for larger resolutions, accurately discerning and characterizing these features becomes challenging. The available spatial resolution impacts the correct characterization of warps, and despite advancements in high-resolution imaging techniques, the small radial scale warps will be susceptible to beam dilution. This limitation increases the complexities involved in observing and characterizing warped disks.

    We investigated the range of inclinations for which the method remains applicable. Specifically, for the warped disk scenario, we verified values of $i_{\text{in}} = (15\degr, 25\degr, 50\degr)$, while for the radial flow case, we tested for an inclination of $i_{\text{out}} = (15\degr, 35\degr, 50\degr)$. All other parameters remained consistent with those outlined in Section \ref{sec:Results}. A summary of the residual maps is shown in Fig. \ref{fig:models_WF_higherInc}. We observed an increase in the amplitude of the residuals with inclination. We suggest that this method works optimally for inclinations below 40\degr.

\section{Conclusion}
\label{sec:Conclusions}

    The prevalence of misalignments challenges the assumption of planar disks, requiring methods capable of accounting these more complex geometries. Molecular line observations indicate that misalignments can mimic signals attributed to radial flows. Through comprehensive warped kinematics and radiative transfer simulations, we have investigated whether this degeneracy can be broken. This study advances the characterization of astrophysical disks, particularly those exhibiting warped structures. Applying this tool to actual observational data, presents an opportunity to revisit and reassess warped systems that may have been misinterpreted, as discussed by \citet{Young2022}. The tool capability to discern between warps and other kinematic features can lead to a more nuanced interpretation of observed structures.
    
    Our main findings can be summarized as follows:

    \begin{enumerate}
        \item Gas kinematics in protoplanetary disks are complex, particularly when warps and radial flows are involved, posing challenges for accurate modeling and interpretation. The \texttt{eddy} modeling tool, extended to include warped geometries and radial flows, provides a valuable framework for investigating these types of disks.
        \item Noise levels comparable to previous ALMA observations do not affect the capability to discern between signatures, because the influence of the inclination $i$ and radial flows on the data surpasses the impact of thermal noise. The accurate characterization of warps in protoplanetary disks is limited by angular resolution.
        \item While degeneracies between different disk configurations, such as warped-only and radial flow models, complicate the interpretation of observed data, signatures that emerge in the residuals from fitting an incorrect model to a system that exhibits twisted kinematics can be used as a criteria to distinguish between the warped disk radial flow dilemma.
        \item Future observational efforts, particularly high-resolution ALMA observations, hold promise for refining our understanding of disk kinematics and evolution. Advancements in modeling techniques and observational capabilities are crucial for overcoming the limitations and complexities associated with characterizing gas kinematics in protoplanetary disks.
    \end{enumerate}

    We also discuss the current limitations that observational studies of warps face. Our methodology establishes a foundation for future investigations and holds promising applications in revisiting and reinterpreting signals from potentially misaligned disks. 

\begin{acknowledgements}

We sincerely thank the anonymous referee for their valuable suggestions, which significantly enhanced the quality of our paper. T.B. acknowledges funding from the European Union under the European Union's Horizon 2020 research and innovation programme under grant agreement No 714769 and under the Horizon Europe Research and Innovation Programme 101124282 (EARLYBIRD) as well as funding by the Deutsche Forschungsgemeinschaft (DFG, German Research Foundation) under grant 325594231, and Germany's Excellence Strategy - EXC-2094 - 390783311. Views and opinions expressed are, however, those of the authors only and do not necessarily reflect those of the European Union or the European Research Council. Neither the European Union nor the granting authority can be held responsible for them.

\texttt{Software:} \texttt{NumPy} \citep{Harris2020}, \texttt{emcee} \citep{Foreman-Mackey2013}, \texttt{Matplotlib} \citep{Hunter2007}. 
      
\end{acknowledgements}

%
%

\bibliographystyle{bibtex/aa} 
\bibliography{main.bib}

\begin{appendix}

\section{Channel maps}
\label{app:channel_maps}
    
\begin{figure*}
    \centering
    \includegraphics[width=17cm]{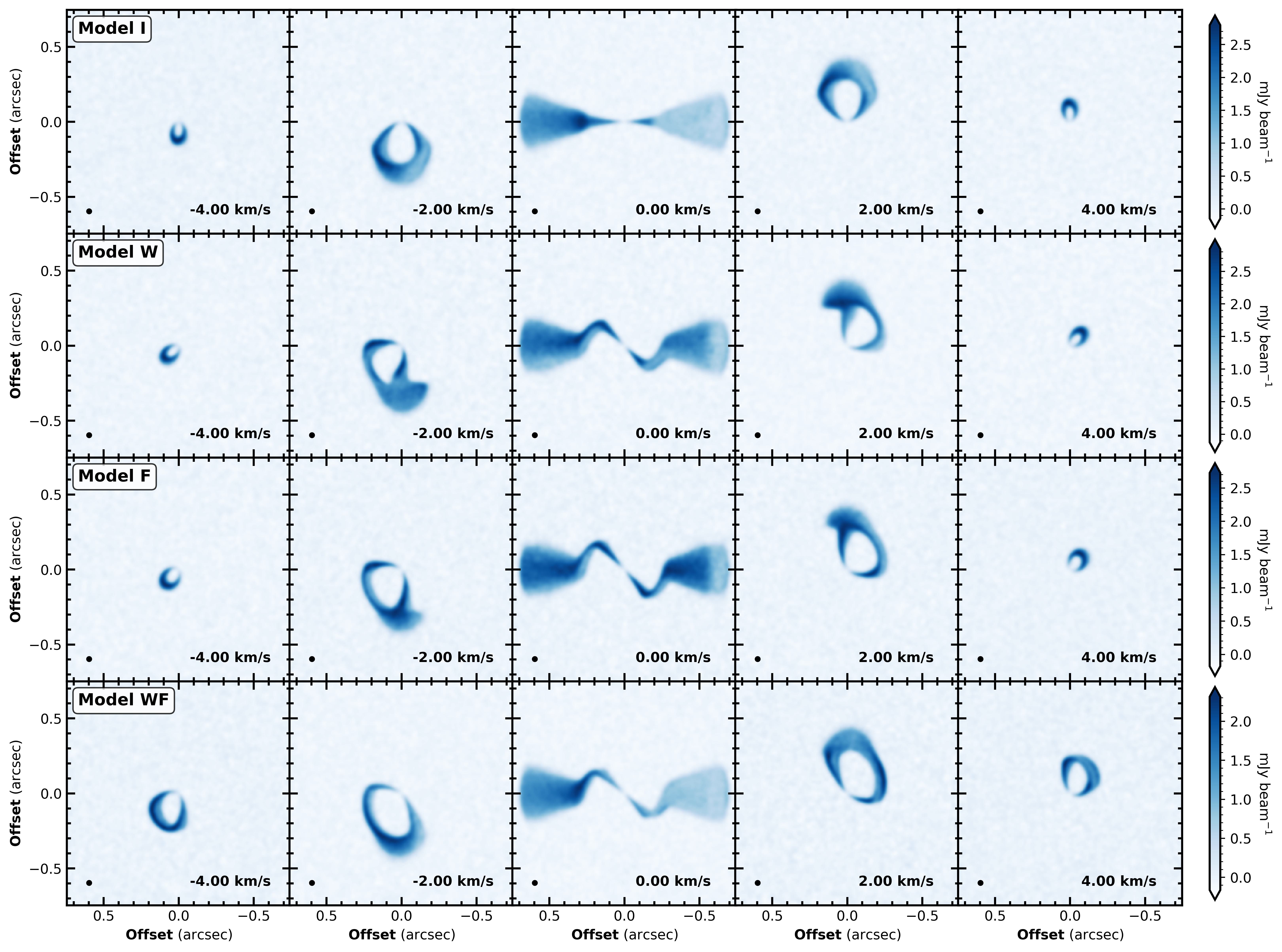}
    \caption{Synthetic CO J=2-1 channel maps, imaged at 40 m s$^{-1}$ spacing for each kind of model. The beam size is shown in the bottom left of the panels. The model type is indicated in the upper left: the top row shows the warped model, middle row shows the unperturbed disk with a radial flow, and the bottom panel shows a disk with a warp and a radial flow. We only show the channels close to the systemic velocity v$_{\text{LSR}} = 0\text{ km s}^{-1}$.}
    \label{fig:channel_warps}
\end{figure*}
\FloatBarrier

\section{Extra figures}
\label{app:incl_noiselevels}

\begin{figure*}
    \centering
    \includegraphics[width=17cm]{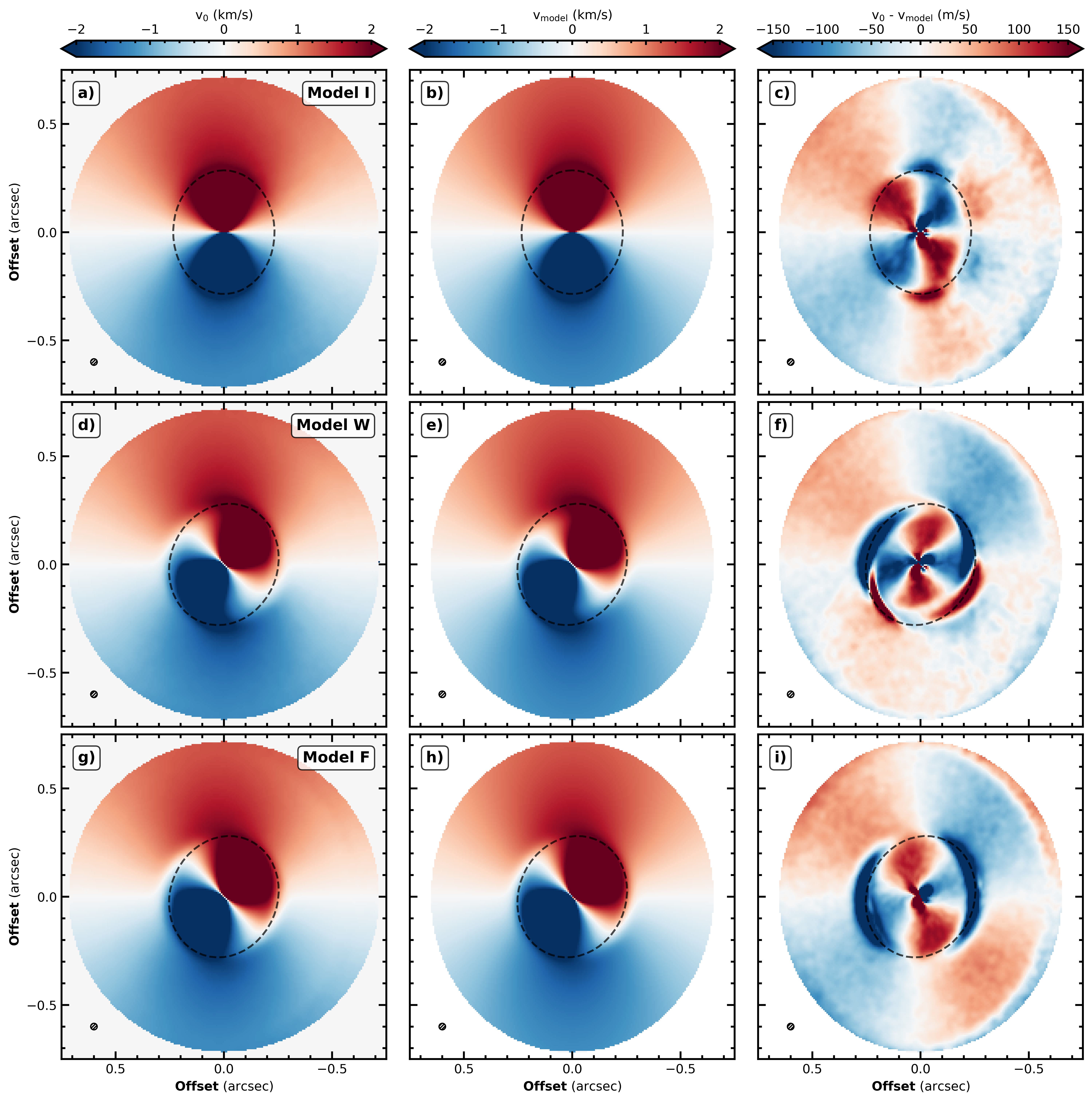}
    \caption{Same as Fig. \ref{fig:models_WF} but with a noise RMS of $\SI{3.5}{\kelvin}$ over $\SI{350}{\metre\per\second}$.}
    \label{fig:models_highNoise}
\end{figure*}

\begin{figure*}
    \centering
    \includegraphics[width=17cm]{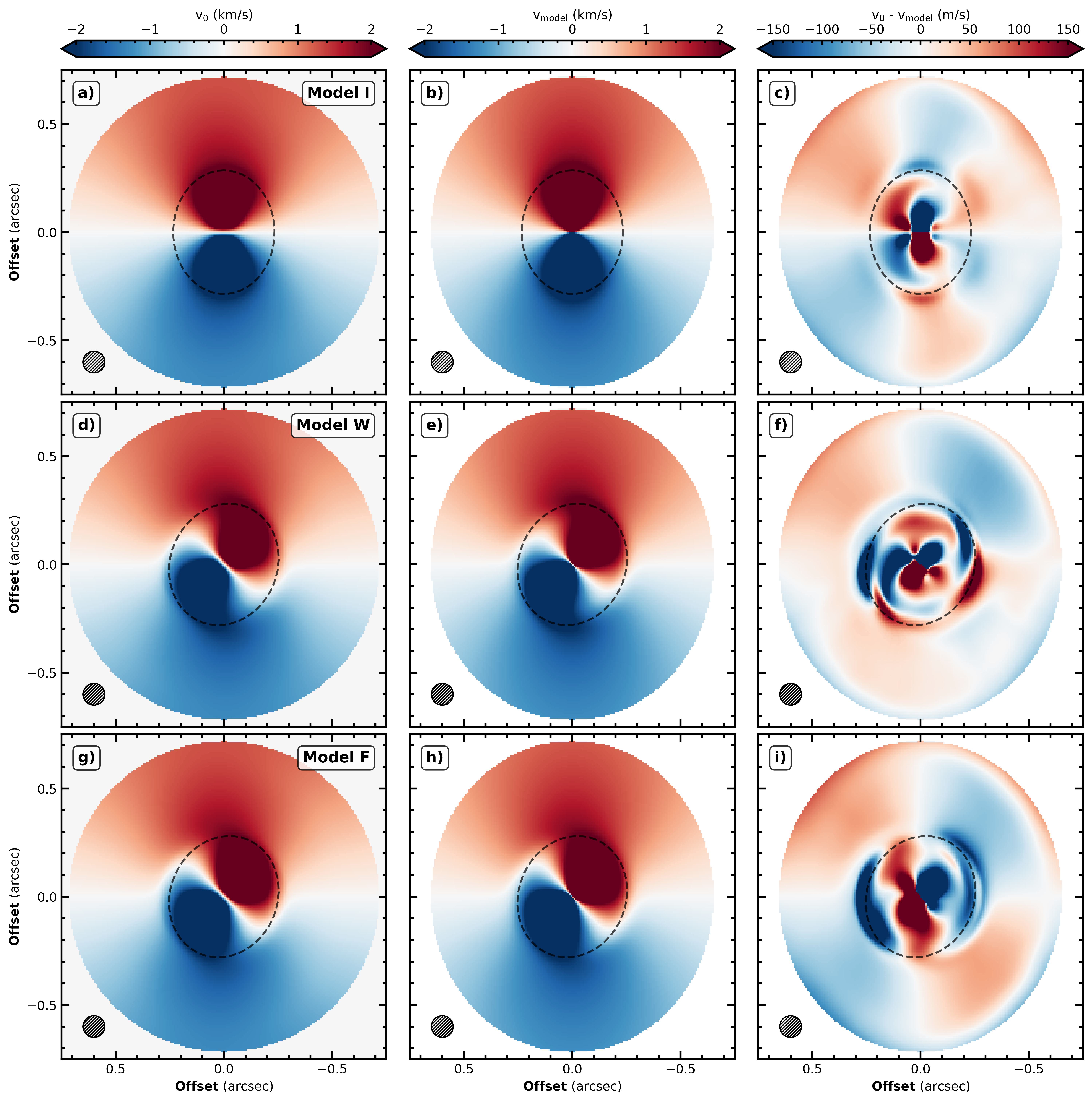}
    \caption{Same as Fig. \ref{fig:models_WF} but the synthetic radiative transfer models are convolved with a 0.1\arcsec\ FWHM beam}
    \label{fig:models_WF_0.1FWHM}
\end{figure*}

\begin{figure*}
    \centering
    \includegraphics[width=17cm]{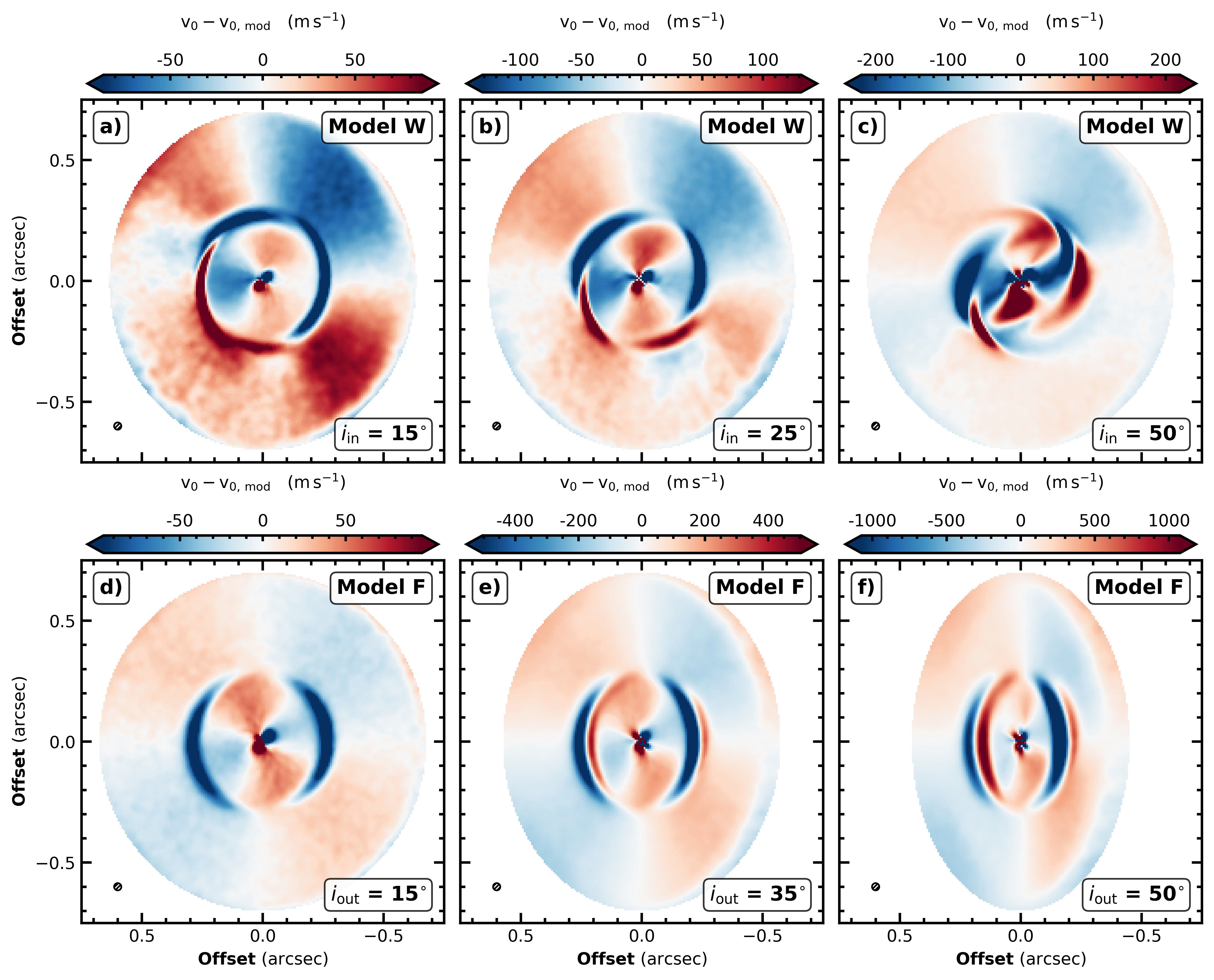}
    \caption{Residual maps after subtraction of their respective best-fit models. The top row shows the residual for Model W with inclinations $i_{\text{in}} = $ (15\degr, 25\degr, 50\degr). Bottom row shows the residual for Model F with inclinations $i_{\text{out}} = $ (15\degr, 35\degr, 50\degr). The respective inclination it shown in the bottom right corner, and the beam size is shown in the bottom left corner.}
    \label{fig:models_WF_higherInc}
\end{figure*}
    
\end{appendix}

\end{document}